\documentstyle[times,pramana,epsf,floats]{ias}

\def\lb{\left(}
\def\rb{\right)}
\def\be {\begin{equation}}
\def\ee {\end{equation}  }
\def\beq{\begin{eqnarray}}
\def\eeq{\end{eqnarray}  }
\def\bi {\begin{itemize} }

\def\ei {\end{itemize}   }
\def\RE {I\kern-6pt R    }
\def\Z  {Z\kern-13pt Z   }
\def\be {\begin{equation}}
\def\ee {\end{equation}  }
\def\beq{\begin{eqnarray}}
\def\eeq{\end{eqnarray}  }

\begin{document}
\mark{{Black Hole Critical Phenomena}{Without Black Holes}}
\title{Black Hole Critical Phenomena Without Black Holes}

\author{Steven L. Liebling}
\address{Theoretical and Computational Studies Group\\
     Southampton College, Long Island University,
     Southampton, NY 11968}

\abstract{
Studying the threshold of black hole formation via numerical
evolution has led to the discovery of fascinating nonlinear
phenomena. Power-law mass scaling, aspects of universality,
and self-similarity have now been found for a large variety
of models. However, questions remain. Here I briefly review
critical phenomena, discuss some recent results, and describe
a model which demonstrates similar phenomena without gravity.
}

\maketitle

\section{Introduction}
The discovery of black hole critical behavior represents an important
example of the power of numerical computation in physics.
The phenomena discovered were completely unpredicted and unexpected
prior to the numerical work of Choptuik~\protect\cite{choptuik}.

In this Proceedings, I briefly review critical phenomena
before describing some research of my colleagues and me.
The field and its understanding have grown quickly, and this
review is necessarily limited in its scope. The choice of
topics  is thus biased towards what I have done and follows closely
my talk. I therefore try to retain some of the informality of
a talk omitting detailed discussion of important topics such
as casting the equations of general relativity as an evolution problem
and the numerical solution of those equations.
Also, a
couple excellent reviews are available~\protect\cite{matt_rev,carsten_rev},
and so I attempt here to give only enough detail
to equip and interest the reader in the research following.
The title of this work is explained in Section~\ref{sec:nlsm}.
I conclude by mentioning some important and general
open questions and directions within the field of critical behavior.

\section{Black Hole Critical Phenomena}
\subsection{The Space of Solutions}
When a sufficiently energetic configuration of matter
finds itself in a small enough
region a black hole forms from the inevitable gravitational
attraction. Lacking sufficient energy, such a configuration
fails to produce a black hole, instead finding some other
end state such as a star or dispersing altogether. Were we
engineers with advanced technology, we might attempt to find
that critical amount of energy necessary to form a black hole.

However, despite some fears to the contrary, such technology
does not exist, so instead we investigate this {\em critical
regime} numerically. The first step is to pick a
matter source to consider.  By this time, quite a variety of
sources have been considered, but here I will start
with the relatively simple case of a single scalar field $\psi$ which
happens to be the first case to have been considered~\protect\cite{choptuik}.

Given some initial condition,
the goal is to be able to determine 
whether or not the configuration forms
a black hole. To do so, we need evolution equations for the scalar
field which come from varying an appropriate action $S$ which couples
the scalar field to gravity
\be
S = \int d^4x \sqrt{-g} \Bigl[ R - \psi_{,\mu} \psi^{,\mu} \Bigr]
\ee
where $\sqrt{-g} d^4x$ is the appropriate volume element, $R$ is
the curvature scalar, and $\psi_{,\mu}$ denotes the partial derivatives
of $\psi$.
The construction of a numerical method with which to evolve
the gravitationally coupled scalar field is certainly non-trivial,
but is described in detail elsewhere.

Being able to determine whether a certain configuration of
scalar field forms a black hole, the next goal is to examine the
configuration space of the model. In other words, we would like
to picture the space of all possible initial configurations of
scalar field and which configurations 
form black holes and which disperse (the only two options in
this model). The problem in picturing such a space is that
it is infinite dimensional (imagine the space formed by specifying
the scalar field at every point in real space).

To circumvent this difficulty,
we can
instead parametrize the scalar field $\psi(r,0)$
at the initial time
by assuming a Gaussian form, so that 
\be
 \psi(r,0) = A e^{ \lb r- R_0 \rb^2 / \delta^2}
\ee
in terms of real constants $A,R_0,\delta$.
The time derivative $\dot \psi(r,0)$ at the initial time must
also be specified, but we can choose to have it vanish.
The configuration space
now has only the three dimensions corresponding to the possible values
of the three constants. However, we go a step further by fixing
two of the constants, say $R_0$ and $\delta$. We now have a family
of initial data with one free parameter $A$. We could consider
other one parameter families, and so we generically call the single
free parameter $p$ (so here $p = A$).

Have we lost something in this simplification of using one parameter
families? It turns out that we have not, which we will see later by considering
a variety of such families.

\begin{figure}[htbp]
\epsfxsize=8cm
\centerline{\epsfbox{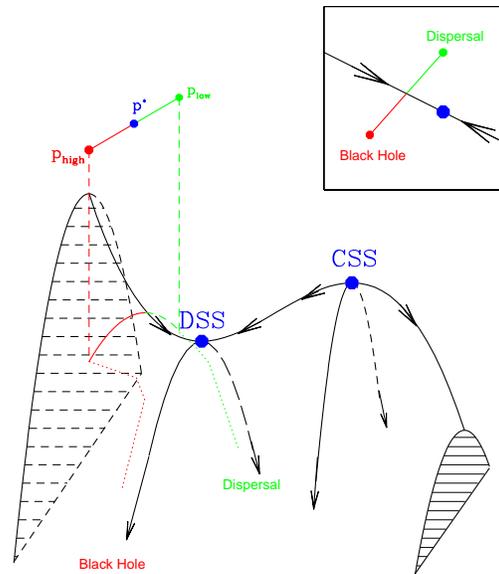}}
\caption{Schematic of Configuration Space. Evolutions which start
         from initial configurations on the near side of the ridge
         form black holes.  Those configurations lying on the
         far side simply disperse.  The critical regime lies on
         the ridge which separates  these two end states.
         The inset shows a bird's eye view looking down on the 
         saddle point.}
\label{fig:phase}
\end{figure}

To complete our map of the configuration space, we need now only determine
for which values of $p$ the configuration evolves into a black hole.
As intuition might suggest, for large values of $p$ black holes do form
and for small values the scalar field disperses. We are now in a position
to picture the configuration space which I sketch in Fig.~\ref{fig:phase}.

Fig.~\ref{fig:phase} shows schematically a section of some generic two-dimensional
({\em i.e.} two parameter)
configuration space. The third dimension (say $z$) represents loosely an
effective potential which describes the fate of the configurations
($x$,$y$). Hence, initial configurations on the near side of the ridge
ultimately evolve to form black holes while those configurations on
the far side disperse. In general, a one parameter family is represented
in this schematic as a line parameterized by $p$ which crosses
the ridge somewhere, an example of which is shown.

This configuration space then reveals two end states (black hole
and dispersal)  which in the jargon of nonlinear dynamics are
called {\em attractors} because trajectories nearby are attracted
to them. These two attractors occur in their respective {\em basins of
attraction} defining the regions in configuration space which are
attracted to each end state. The critical region of this configuration
space is then the boundary between these two end states ({\em i.e.}
the ridge in the sketch of Fig.~\ref{fig:phase}). One aspect of this
schematic to keep in mind is that evolutions do not stay in this simplified
space of initial configurations; the schematic is meant only to help picture
general features of initial data, namely to what end state they evolve.

Stepping back from the abstractness for a moment, consider the actual
process of studying the critical regime. A family of initial data is
chosen and then an initial value of the parameter $p$. This initial data
is then evolved and its final state determined. Say a black hole forms.
Then this value of the parameter becomes the high value $p_{\rm
high}$ and a smaller value of $p$ is chosen. Say the evolution for this
$p$ corresponds to dispersal. Then this value becomes the low
value $p_{\rm low}$. The interval between these two values
 $[p_{\rm low}, p_{\rm high}]$ brackets the critical regime.
To narrow the bracket, one uses a bisection search in which one evolves
the average of the current bracket $p=\lb p_{\rm low}+p_{\rm high} \rb /2$
and updates the bracket according to the evolution.
The iteration of such a search eventually finds a bracket in which
the high and low values approach the
computer's ability to differentiate them. At this point
$p_{\rm low} \approx p_{\rm high}$ and in between them somewhere is
the exactly critical value of $p$ called $p^*$.

The above search essentially explores the region of configuration space
near the ridge. What happens in this regime is the subject of black
hole critical phenomena.

\subsection{Phenomena Occurring in the Critical Regime}
As the search is continued, a certain solution,
the {\em critical solution}, corresponding to $p\rightarrow p^*$
is approached. Evolutions for $p$ near $p^*$ resemble the critical
solution for some time, but eventually tear away from it and head
for either black hole formation or dispersal.
Imagine trying to balance 
a coin on its edge. The better balanced the coin becomes,
the longer the coin ``remains near'' to the critical solution,
that being the static state critically balanced between falling left
and falling right.

In many of the cases studied, the critical solution
demonstrates a type of self-similarity. Consider a general field configuration,
say $\phi(r,t)$. However, let us change coordinates using $u\equiv \ln r - \ln |t|$ 
and $v \equiv \ln r + \ln |t|$ where the collapse time is chosen to occur at $t=0$.
Solutions $\phi(u,v)$ which can be written as
functions of $u$ alone
\be
\phi(u,v) = \phi\lb u\rb
\ee
are {\em self-similar}. Such self-similar solutions are invariant
under changes of scale accompanied by an appropriate change in time.

Remarkably, in the free scalar field case, the critical solution
exhibits a discrete form of such a self-similarity, called 
{\em discretely self-similar} (DSS), in contrast to the above
case called {\em continuously self-similar} (CSS). A solution
which exhibits discrete self-similarity is no longer independent of $v$
but instead periodic in $v$
\be
\phi\lb \ln r, \ln |t| \rb = \phi \lb \ln r \pm n \Delta, \ln |t| \pm n \Delta \rb,
\ee
with periodicity given by the dimensionless number $\Delta$.
The DSS found by Choptuik for the real scalar field has $\Delta=3.44$.
The nature of this discrete self-similarity is still a bit of a mystery,
but, loosely speaking the CSS can be considered some generalization of
a {\em fixed point} to the infinite dimensional case while the DSS
would correspond to a generalized {\em limit cycle}.

I speak here of {\em the} critical solution for the scalar field,
but it is not {\em a priori} clear that there should be only
one solution. In fact, one might guess that the family one
picks will determine the critical solution one
observes. That this is not the case demonstrates the {\em universality}
of the critical solution. That is, independent of what family one
picks, the critical search will always approach this same self-similar
critical solution. 

This universality is now understood in terms of perturbation
theory~\cite{koike}. Consider linear perturbations to
the exactly critical solution existing
inside the full infinite-dimensional phase space of the model.
The critical solution will have a number of modes, stable
or unstable. The stable modes decay, and hence drive the perturbed
solution back to the critical solution. However, the unstable modes
are perhaps more interesting, driving the solution away from
criticality.

The key point here is that the critical solution can have only
one unstable mode. It is the action of this unstable mode near
the critical point which drives slightly sub-critical solutions
to dispersal and super-critical solutions to black hole formation.
Hence, the one parameter tunes the excitation of this
single unstable mode while the stable modes drive the solution
towards the critical solution. Such a solution is an
{\em intermediate attractor} because it is only one unstable
mode away from being an attractor. Because of these dynamics,
tuning similar families of initial data are driven (modulo
the unstable mode) towards the {\em same} critical solution.

This picture of the perturbative modes also explains
another feature of critical behavior, namely {\em
power law mass scaling} of the black holes formed.
As stated above, tuning of $p$ drives the excitation of the
unstable mode to zero, and hence for slightly supercritical
evolutions, the scale of the excitation is set by $p-p^*$.
It can then be shown that the mass of the black holes formed
by such evolutions follows a power law
as
\be
M \propto \lb p - p^* \rb^\gamma
\ee
for some constant $\gamma$ determined by the model.
In the case of the free scalar field, the mass scaling exponent
about the DSS is $\gamma = 0.37$.
The perturbative picture has successfully shown that $\gamma$
is given by the inverse of the eigenvalue of the single
exponentially growing mode of the critical solution.

We consider how these solutions appear at times
near to collapse. Whether CSS or DSS, the dynamics appear
at smaller and smaller spatial scales which correspond to regions
of arbitrarily high curvature visibly at infinity.
The exactly critical solution has a naked singularity.
Critical behavior thus becomes a way to find naked singularities.

The discussion above applies directly to regions of phase space
in which a self-similar solution plays the role of the critical
solution. In these cases, black holes form with arbitrarily small
mass (for $p$ arbitrarily close to $p^*$), and in analogy to phase
transitions with order parameter $M$, these case are called {\em Type~II}.
Other models not discussed here possess stationary solutions which
act as critical solutions in which case the black hole formation
has a finite lower limit for their mass. Such models exhibit Type~I collapse.

\section{Two, Three, and Four Fields}
We have considered so far the simple case of just one scalar field.
By now a large number of different matter models have been considered
with a variety of interesting twists and turns in the picture presented
above. Here I describe a few models I have studied which happen to 
categorize themselves by the number of scalar fields they employ.
For background, I make note of a few useful resources which discuss
such multiple scalar field models. Ryder discusses the real and complex
scalar fields as a precursor to quantum field theory~\cite{ryder}.
Rajaraman discusses monopoles and charge conservation from a 
mathematical perspective~\cite{rajaraman} while Vilenkin and Shellard
present the standard discussion of these models in the context of
topological defects~\cite{vilenkin}.

\subsection{Two Fields: Generalized Complex Scalar Field}
Consider a complex scalar field $F(r,t)\equiv \psi(r,t) + i \phi(r,t)$ 
in terms of real scalar fields $\psi$ and $\phi$ with the action
\begin{equation}
S= \int d^4x \sqrt{-g} \left(
      R -\frac{2|\nabla F |^2}{\left( 1 - \kappa |F|^2 \right)^2 }
   \right),
\label{eq:action}
\end{equation}
where $\kappa$ is a dimensionless
parameter of the model. For $\kappa = 0$, the model reduces to a
free complex scalar field while for general $\kappa$ the model
is the nonlinear sigma model with target manifold corresponding
to surfaces of constant curvature $-\kappa$. For negative values of
$\kappa$ the target space is $S^2$ corresponding to the group
$SO(2)$~\cite{lieblingmultiply}.

For $\kappa=0$ the model corresponds to the free complex scalar field
for which the $\Delta = 3.44$ DSS is the known
attracting critical solution.
However, 
Hirschmann and Eardley study this model and find that if $\kappa$
is increased above $\kappa \approx 0.075$ a bifurcation occurs
such that a CSS becomes the attracting critical solution~\cite{hirschmann}.
This bifurcation occurs in a region of parameter space equivalent to
that of a free scalar field coupled to Brans-Dicke gravity, and the
bifurcation has been confirmed in that model~\cite{liebling_brans}.

\begin{figure}[htbp]
\epsfxsize=8cm
\centerline{\epsfbox{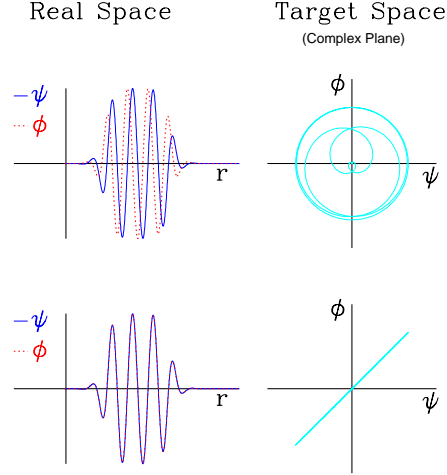}}
\caption{Aspects of the spiral initial data. On top is the
         spiral data Eq.~(\ref{eq:spiral}) for $\Delta \varphi =0$ and
         on bottom is that for $\Delta \varphi =-\pi/2$. On the left
         are shown the real and imaginary parts of the complex
         field $F(r,0)$ as functions of space. On the right
         is shown $F(r,0)$ in the complex plane.
         }
\label{fig:tuning}
\end{figure}

What a surprise then to find that the above results fail to hold
with a certain family of initial data that I call {\em spiral} data.
I consider initial data for $F(r,t)$ such that the real and imaginary
parts are
\beq
\psi(r,0) & = & p~f(r) \cos (\omega r + \Delta \varphi) 
\nonumber \\
\phi(r,0) & = & p~f(r) \sin (\omega r ),
\label{eq:spiral}
\eeq
where $p$ and $\Delta \varphi $ are real constants and the function $f(r)$
serves as an envelope smoothly approaching zero for large
and small $r$. For simplicity of some of these arguments,
I assume that the initial time derivatives of the fields vanish.

Choosing Eq.~(\ref{eq:spiral}) with $\Delta \varphi =0$ and
tuning $p$ finds the CSS critical solution for $\kappa=0$
despite the fact that the CSS has multiple unstable modes (and
hence is not an intermediate attractor) for $\kappa \alt 0.075$.
Similarly, fixing $\Delta \varphi =-\pi/2$ and tuning $p$ finds the DSS
critical solution for $\kappa\agt 0.075$ despite it not being
an intermediate attractor.

To understand this behavior which would appear to go against the
grain of the understanding presented above, we first consider 
certain aspects of this initial data as shown in Fig.~\ref{fig:tuning}.
For $\Delta \varphi =-\pi/2$, we have $\psi(r,0)=\phi(r,0)$ which corresponds
to having no charge in the initial data. That this data has no charge
can be seen in two ways. First, the model has a global $U(1)$
symmetry under which the model is invariant to global rotations
of $F(r,t)$ in the complex plane. Thus, we can rotate $F(r,0)$
until it becomes all real. In this case, the model is equivalent to
a single scalar field which has no charge (two fields are necessary
for there to be charge). Secondly, one can assess the charge of this
initial data simply by looking at the area spanned in the complex plane.
As shown in Fig.~\ref{fig:tuning}, for $\Delta \varphi =-\pi/2$ no area is
spanned.

Contrast this case with the case of $\Delta \varphi =0$. In this case, the field
can be expressed as $F(r,0) = f(r) \exp\lb i \omega r\rb$. In regions
where the envelope function $f(r)=1$, the field configuration
maximizes the area spanned in the complex plane, and hence maximizes
the charge density for a given energy density~\cite{lieblingmultiply}.

\begin{figure}[htbp]
\epsfxsize=8cm
\centerline{\epsfbox{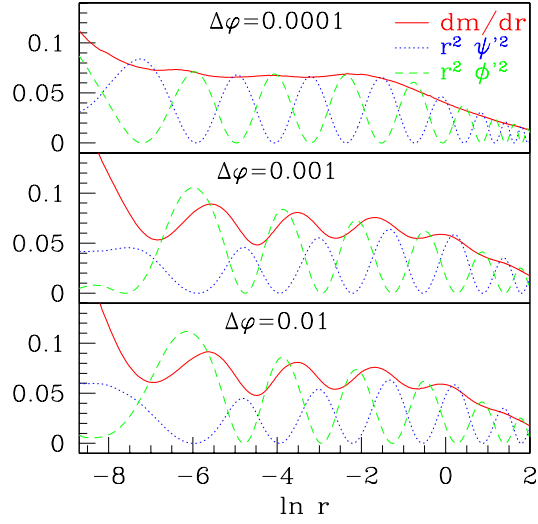}}
\caption{The critical solution found by tuning $p$ in initial data
         of the form Eq.~(\ref{eq:spiral}) for $\kappa=0$
         with different values of
         $\Delta \varphi$. For $\Delta \varphi \rightarrow 0$,
         the CSS critical solution is observed where, for generic
         initial data, the DSS would be observed. As $\Delta \varphi$
         is increased, however, the critical solution approaches the
         DSS.
         }
\label{fig:spiral}
\end{figure}

Fig.~\ref{fig:spiral} demonstrates the case of perturbing $\Delta \varphi$
away from zero. While for $\kappa=0$ with $\Delta \varphi=0$ one finds
the CSS solution where one would otherwise expect to find the DSS,
as $\Delta \varphi$ is perturbed, tuning $p$ quickly moves to the
DSS. This indicates that the case of $\Delta \varphi=0$ is indeed quite
special.

These observations suggest that changes to $\Delta \varphi$ represent
tuning a second parameter. While tuning $p$ effectively tunes the mass
of any black hole formed, tuning $\Delta \varphi$ tunes the charge.
Hence, while the CSS has other unstable modes (a pair with complex conjugate
eigenvalues) for $\kappa=0$, the spiral initial data with $\Delta \varphi=0$
represents data already tuned for maximal charge.
This tuning disallows critical solutions with no charge such as the DSS.

\subsection{Three Fields: Global Monopoles}

Consider a triplet scalar field $\Phi^a$ ($a=1,2,3$) coupled to
gravity with the Lagrangian
\be
L =
     - \frac{1}{2} \Phi^a{}^{;\mu} \Phi^a{}_{;\mu}
     - \frac{1}{4}\lambda \left[ \Big( \Phi^a \Big)^2 - \eta^2 \right]^2
\label{eq:lagrange}
\ee
where $\lambda$ is a constant representing the coupling of the potential
and $\eta$ is the scale of symmetry breaking.
Also assume  the hedgehog ansatz for the triplet field
\be
\Phi^a(r,\theta,\varphi,t) = f(r,t) {\hat r}^a,
\ee
with which  we need only specify and evolve the field $f(r,t)$.

One interesting aspect to this model is the presence of static
solutions $f_s(r)$, namely global monopoles. Such solutions are found
with a usual shooting method by which one adjusts the value
$a \equiv f_a'(0)$, integrates outward to large radius, and shoots
for an asymptotic boundary condition such that
$f_s(r\rightarrow \infty) \rightarrow \eta$. Such solutions represent
global monopoles of unit charge and are parametrized by $a(\eta)$.

\begin{figure}[htbp]
\epsfxsize=8cm
\centerline{\epsfbox{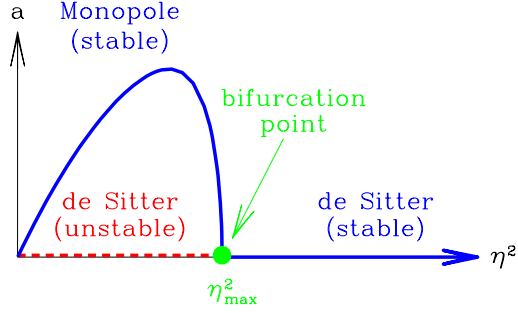}}
\vspace{-1.0in}
\caption{Schematic of static global monopole solutions in terms of
         scale of symmetry breaking $\eta$ and $a$, where $a\equiv f_s'(0)$.
         The static monopole solutions bifurcate with de Sitter space
         at $\eta_{\rm max}$ and stability is exchanged.
         }
\label{fig:fig3}
\end{figure}

It turns out that above a critical value $\eta=\eta_{\rm max}$ no such solutions
can be found~\cite{liebling_static}. A diagram of these solutions
in terms of $\eta^2$ is
shown in Fig.~\ref{fig:fig3}.
As $\eta$ is increased towards the critical $\eta_{\rm max}$, the value
of $a$ for which the static monopole solution is found decreases toward
zero. This approach to zero represents the point at which
the static monopole becomes identical to de Sitter space in which
$f_s(r)=0$ and the symmetry-breaking potential reduces to a cosmological
constant~\cite{maisonliebling}.

Within this model, one can consider evolving generic families of initial
data specifying the scalar function $f(r,t)$.  Consider first
Gaussian families such as
\be
 f(r,0) = A e^{ \lb r- R_0 \rb^2 / \delta^2}
\ee
in which no monopole charge is present.
With such a family, a DSS solution is 
observed as the critical solution, though this is not the same
DSS as observed by Choptuik. Instead, I find a DSS with $\Delta = 0.46$
and $\gamma = 0.119$~\cite{liebling}.

\begin{figure}[htbp]
\epsfxsize=8cm
\centerline{\epsfbox{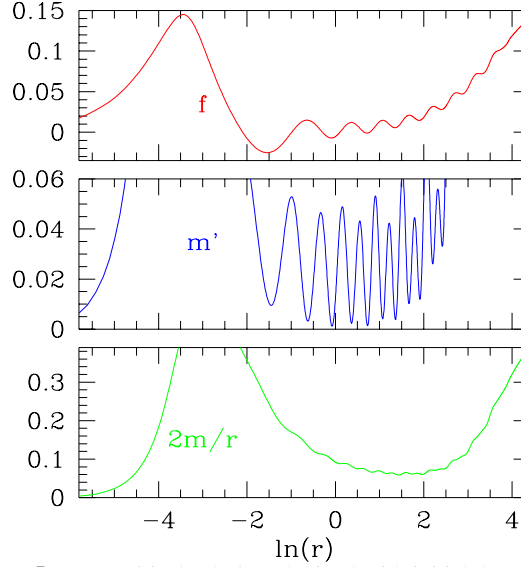}}
\caption{Near critical solution obtained with initial data of type
         Eq.~(\ref{eq:mono_init}). Shown is the function $f(r,t)$,
         the mass density $m'(r,t)$, and the value of $2m(r,t)/r$
         as functions of $\ln r$. In the top frame, one can see
         the approach of $f(r,t)$ to its background large-$r$ value
         $\eta = 0.15$.
         }
\label{fig:mono}
\end{figure}

However, a more interesting family is a Gaussian
pulse as a (nonlinear) perturbation to the static monopole solution
\be
 f(r,0) = f_s(r) + A e^{ \lb r- R_0 \rb^2 / \delta^2}.
\label{eq:mono_init}
\ee
Observations of these families show the pulse traveling without
disturbing the monopole solution. That is, the monopole persists
as the pulse evolves demonstrating its stability. This stability
was then explicitly confirmed via a linear analysis~\cite{maisonliebling}.
Because the static monopole solutions are stable, no Type~I
critical behavior is expected nor observed. Instead, with the background
provided by the global monopole, the usual Type~II critical behavior
is observed as shown in Fig.~\ref{fig:mono}. The figure shows a
DSS critical solution with $\Delta=0.46$ on a global monopole
background. Hence, the potential is
asymptotically irrelevant, as discussed in~\cite{review2}.

Maison has extended this work and found families of excited, static 
monopoles with multiple zeros of the Higgs field $f(r)$~\cite{maison}. These
solutions appear as separate arcs (within the one shown)
on the solution space schematic of Fig.~\ref{fig:fig3}
These solutions have an increasing number of unstable modes such
that the first excited family has one unstable mode. Hence, this
first excited family might occur in the context of Type~I critical
collapse, though
such solutions have small cosmological horizons making meaningful evolutions
difficult~\cite{maison_conjecture}.

\subsection{Four Fields: Nonlinear Sigma Model}
\label{sec:nlsm}
We have now looked at two scalar fields (the complex scalar field) and
three fields (the global monopole), and now turn to four
fields. Here again we
consider a component scalar field $\Phi^a$
with the hedgehog ansatz, however now $a$ runs over $1,2,3,4$.
With the hedgehog ansatz, the quartic field is
\be
\Phi^a = \left( \begin{array}{l}
                \sin \chi(r,t) \sin \theta \sin \varphi \\
                \sin \chi(r,t) \sin \theta \cos \varphi \\
                \sin \chi(r,t) \cos \theta \\
                \cos \chi(r,t)
                \end{array} \right)
\label{eq:hedge}
\ee
in terms of another real scalar field $\chi(r,t)$.
This field satisfies the same Lagrangian as the monopole case,
Eq.~(\ref{eq:lagrange}), with the modifications that $a=1,2,3,4$
and gravity is not coupled to the field.
We also work in the nonlinear sigma model approximation in
which $\lambda$ is treated as a Lagrangian multiplier, and hence
the field pays an infinite price in energy to leave the vacuum
manifold. Because of this infinite price, the field is restricted
to the vacuum manifold ($S^3$).
That the ansatz
Eq.~(\ref{eq:hedge}) is everywhere and for all time on the
vacuum manifold can be seen by observing that $|\Phi^a| = 1$
(we have the freedom to set $\eta=1$ retaining full generality).

The equation of motion for $\chi(r,t)$ is then
\be
\ddot \chi
          - \frac{1}{r^2} \left( r^2 \chi' \right)'
          = - \frac{\sin \left(2\chi\right)}{r^2},
\ee
where a dot represents the partial derivative with respect to
$t$ and a prime denotes the partial derivative with respect to 
$r$.

This model is not coupled to gravity and so no black holes can
be formed. However, mathematicians have shown for this model that some sets of
regular initial data develop singularities within some finite
time~\cite{singular} while other sets of initial data  are regular
for all time~\cite{small data1,small data2}. Thus the model
has two end states, dispersal and singularity formation, and
presumably there is some critical regime separating the two.

To investigate this region, we once again choose a one parameter
family such as a Gaussian
\be
 \chi(r,0) = A e^{ \lb r- R_0 \rb^2 / \delta^2},
\label{eq:gauss_nlsm}
\ee
and consider evolutions for different $p=A$. Indeed, for small
$p$ we observe the initial Gaussian pulse evolving towards large
radius with no evidence of singularity formation. In other words,
for small $p$, the initial data simply disperses. For large $p$
we observe unbounded growth of $\chi'$ and $\dot \chi$ at the origin
signaling singularity formation.

Bracketing the critical regime $p^*$ for the family in
Eq~(\ref{eq:gauss_nlsm}), we approach a continuously self-similar
solution for $\chi(r,t)$. Further for a few families of initial data
with compact support, the same self-similar solution is approached
indicating that the critical solution is universal within some region
of parameter space~\cite{isenberg,bizon1}.

\begin{figure}[htbp]
\epsfxsize=8cm
\centerline{\epsfbox{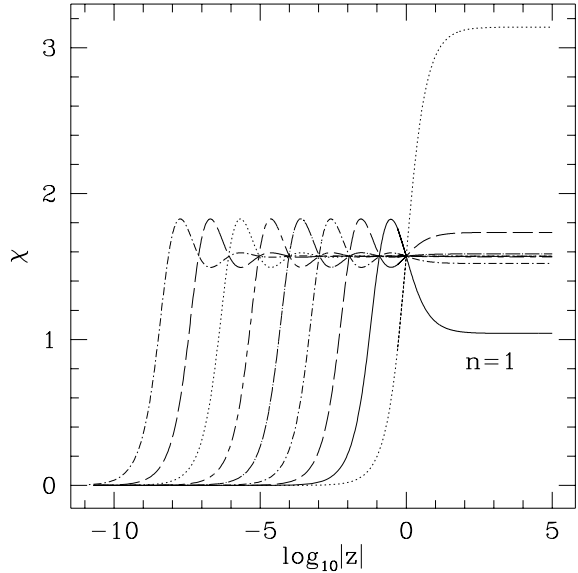}}
\vspace{0.2in}
\caption{Self-similar solutions for $\chi(r,t)$ in the nonlinear
         sigma model.
         The label $n$ labels the number of times $\chi$ crosses
         the line $\pi/2$ on the interval $(0,1)$ where  $z\equiv -r/t$.
         The $n=1$ solution is the critical solution for initial
         data families of compact support.
         }
\label{fig:AB_solns}
\end{figure}

It turns out that the model permits a family of self-similar solutions
as found by~\cite{aminneborg,bizon2}.
The first nine of these solutions are shown
in Fig.~\ref{fig:AB_solns}. These solutions are parameterized by the
number of times $\chi$ crosses $\pi/2$ between $z=0$ and $z=1$ where
$z\equiv -r/t$. However, the $n=0$ solution has no unstable modes, and the
other solutions become increasingly unstable,
appearing to have  $n$ unstable spherically symmetric modes.
The $n=1$ solution, having just one unstable mode,
appears to be the intermediate attractor.

Sprinkled in the discussion above have been qualifications about
initial data of compact support. What if we remove such a restriction?
We could then consider the collapse of so-called {\em textures} which
are necessarily non-compact requiring $\chi$ to asymptote to integer
multiples of $\pi$~\cite{vilenkin}. Collapse of such global initial
data is discussed in~\cite{isenberg}, but it is not clear what behavior
is contained in the critical regime.

That such a non-gravitational model shares some of the same
phenomena as the black hole case suggests that many other 
models might contain interesting critical behavior.

\section{Questions For the Future}

\subsection{The Future: Less symmetry}
The work described above all assumes spherical symmetry.
In fact, the only numerical evolutions geared towards examining
critical behavior conducted in a less
restrictive symmetry than spherical symmetry has been that
of Abrahams and Evans~\cite{abrahams}. They modeled the collapse
of axisymmetric gravitational waves. Moving from spherical
symmetry to axisymmetry requires a large increase in
numerical difficulty and sophistication which makes their work
quite remarkable.

There has also been important work away from spherical symmetry using
perturbation theory. The DSS critical solution for the real scalar
field has been shown to have only decaying non-spherical modes,
and hence is expected to be found in general collapse for initial
data near spherical symmetry~\cite{carsten_choptuik}. Perturbation
theory has also found a scaling relationship for small angular momentum
of black holes formed near criticality~\cite{carsten_ang}.

Much work remains though in higher dimensions. Building on the work
of Abrahams and Evans, evolutions of scalar fields and other
matter sources in axisymmetry
might find new, non-spherically symmetric critical solutions.
The scaling of angular momentum for large values can be investigated
as can the stability of the known critical solutions to non-spherically
symmetric modes.
Because of such promise,
Matt Choptuik, Eric Hirschmann and I are currently constructing
an axisymmetric evolution
code.

\subsection{Questions}
In addition to numerical work in higher dimensions, many other questions
remain. That a discretely self-similar solution occurs at criticality
adds to the interest of critical phenomena, but it is not clear
why such a symmetry is found there. Garfinkle's work is one of
the few to start down this path, examining Einstein's equations
in terms of both a scale invariant part and an overall scale~\cite{garfinkle}.
Hopefully, more work in this direction might reveal from where
the actual value of $\Delta$ is derived and what dependence it
has on the matter source.

In this same vein, one wonders at the similarity between the
CSS of Hirschmann and Eardley and the DSS. This particular CSS
solution is self-similar, but retains a periodic nature similar
to the DSS by way of a rotation in the complex plane. In other
words, quantities invariant to global $U(1)$ rotations
in the complex plane such as the energy density are perfectly
self-similar, while phase-dependent fields such as the components
of the complex field $F(r,t)$ are discretely self-similar.
The energy density associated with the
real or the imaginary components oscillates with changes
in scale (this effect is apparent in Fig.~\ref{fig:spiral}).
In the DSS however, these oscillating energy
densities are in phase, and hence the solution as a whole
is DSS, not CSS.
Such a similarity would seem to indicate something rather
deep in the equations.

Finally, how generic is critical behavior? Similar behavior
is seen as described above in a model with no gravity, the
nonlinear sigma model. Presumably there are any number of
models with various attracting end states between which lies
a critical surface yet to be discovered. What other symmetries
might these critical solutions have? Might there be critical phenomena
studied experimentally instead of numerically just as topological
defects can be studied in liquid crystals?

\section{Acknowledgments}
I would like to thank my collaborators
M.W.~Choptuik,
E.W.~Hirschmann,
J.~Isenberg,
and
D.~Maison. I would also like to acknowledge
the support of NSF PHY-9900644.

\clearpage


\end{document}